\documentclass[aps,pre,twocolumn,superscriptaddress,showpacs,draft]{revtex4}
\input{epsf}
\usepackage{color}
\usepackage{amsmath}
\usepackage{amsfonts}
\usepackage{dcolumn}                             
\usepackage{amssymb}
\usepackage{graphicx,graphics,wrapfig,rotating}         
\usepackage[german,english]{babel}
\usepackage{bm,fancybox}                         
\usepackage{bbm}

\newcommand{\rem}[1]{}

\newcommand{\ket}[1]{|\,#1\,\rangle}                %
\newcommand{\bra}[1]{\langle\,#1\,}                 %
\newfont{\Bb}{msbm10}                   %
\newcommand{\tr}{\mbox{tr}}             %

\def\bra#1{{\langle#1|}}
\def\ket#1{{|#1\rangle}}

\def\tr{{\rm Tr}}

\begin{document}

\title{On the environmental stability of quantum chaotic ratchets}

\author{Gabriel G. Carlo}
\affiliation{Departamento de F\'\i sica, CNEA, Libertador 8250, (C1429BNP) Buenos Aires, Argentina}
\author{Leonardo Ermann}
\affiliation{Laboratoire de Physique Th\'eorique, UMR 5152 du CNRS, Universit\'e Paul Sabatier,
31062 Toulouse Cedex 4, France}
\author{F. Borondo}
\affiliation{Departamento Qu\'\i mica, and Instituto Mixto de
Ciencias Matem\'aticas CSIC - UAM - UC3M - UCM, Universidad
Aut\'onoma de Madrid, Cantoblanco, 28049 Madrid, Spain}
\author{R. M. Benito}
\affiliation{Grupo de Sistemas Complejos, and Departamento de F\'\i sica y Mec\'anica,
Escuela T\'ecnica Superior de Ingenieros Agr\'onomos,Universidad Polit\'ecnica de Madrid,
28040 Madrid, Spain}

\date{\today}


\begin{abstract}

The transitory and stationary behavior of a quantum chaotic ratchet consisting of a
biharmonic potential under the effect of different drivings in contact with a thermal
environment is studied.
For weak forcing and finite $\hbar$, we identify a strong dependence of the
current on the structure of the chaotic region.
Moreover, we have determined the robustness of the current 
against thermal fluctuations in the very weak coupling regime.
In the case of strong forcing, the current is determined by the shape of a chaotic attractor.
In both cases the temperature quickly stabilizes the ratchet,
but in the latter it also destroys the asymmetry responsible for the current generation.
Finally, applications to isomerization reactions are discussed.

\end{abstract}

\maketitle

\section{Introduction}
  \label{sec:intro}

Directed transport, understood as transport phenomena in periodic systems out of equilibrium,
has attracted much attention in recent years \cite{Feynman,Reimann}.
As a consequence the field has developed into a well established area of statistical physics,
which involves many interdisciplinary aspects.
In this respect, many classical questions and very recently many quantum
issues have been answered.
The breaking of all spatiotemporal symmetries leading to momentum inversion
has been found to be the general mechanism to engineer ratchet systems \cite{origin}.
For the Hamiltonian case, an efficient sum rule explaining the values of the resulting 
net current has been devised \cite{Ham}.
Quantum effects were considered to analyze the first so-called quantum ratchets \cite{Qeffects},
while recently purely quantum ratchets have been found to exist \cite{qratchets}.
Floquet theory has provided with a general explanation for the appearance of a quantum
current in periodic systems \cite{asymFloquet}.

Ratchet systems are interesting in a very broad range of situations,
such as in applications to molecular motors in biology \cite{biology} or to
nanotechnology \cite{nanodevices}.
Cold atoms in optical lattices are one of the main examples of successful
implementations and theoretical developments \cite{CAexp,AOKR}. 
Also, Bose-Einstein condensates have been transported (for particular initial conditions)
by using purely quantum ratchet accelerators \cite{BECratchets}.
In this case, the current has no classical counterpart \cite{purelyQR},
and the energy grows ballistically \cite{recentStudies,coherentControl}.
A possible application field that has not been much explored yet is represented by 
molecular processes, like for example isomerization.
This particular type of chemical reactions have a tremendous relevance in
important biological processes, such as human vision \cite{isovision}
or proton transfer \cite{isoproton}, and their control has also been
considered in the literature \cite{isocontrol}.
Actually, some of us have very recently proposed a novel method to perform
this control in a realistic model for LiNC$\leftrightharpoons$LiCN \cite{isocontrolus},
which is an adequate prototypical example for this kind of processes.
Notice that in these studies the main ingredients used in the present paper,
namely dissipative dynamics associated with thermal noise
and external perturbations \cite{isomerization}, are included. 
However, they were not formulated and studied from the directed transport perspective.
Here we focus on the behavior of a dissipative ratchet in regimes that can also
be of interest for this kind of experiments.

In this paper we study the influence of the environment on a quantum chaotic ratchet.
Our model consists of a mass particle in a biharmonic potential subjected
to different periodic driving forces.
In the dissipative case, the environment can be directly responsible 
for the transport generation.
Let us remark that the results for this regime are still
scarce since the calculations are difficult to carry out,
and then many questions remain still unsolved \cite{QChDissRat,QChDissRat2,ThEff}.
New results seem to provide some solutions to these problems \cite{qflucs,weaklevy}.
Among them, determining the stability of the current is of great relevance and
has recently raised high interest \cite{stability}.
We show that for weak forcing and finite $\hbar$ values
the current strongly depends on the structure of the chaotic region.
Thermalization brings stability, without completely washing out the asymmetric
structures that give rise to it.
In the strong forcing scenario, the current is explained by the asymmetry of a
chaotic attractor.
A stable classical and quantum current is achieved at short times, and the effect
of moderate temperature consists of making this times even shorter,
but at the price of diminishing the current value.

The organization of this paper is as follows.
In Section \ref{sec:model} we present our model for the system and the environment,
also describing the methods used to investigate
the current behavior.
In Section \ref{sec:results} we show the results, and the roles of the coupling
strength and the temperature are analyzed in detail.
Finally, in Section \ref{sec:summary} we summarize our conclusions.

\section{Model and methods}
  \label{sec:model}

Our system consists of a particle moving in a time dependent potential given by
%
\begin{eqnarray}
  V(x,t) & = & 1- \cos(x)- A \cos(2x+\phi_a) +           \\ \nonumber
         &   & F \sin(x) \: [\cos(t)+ B \cos(2t+\phi_b)],
\end{eqnarray}
where $F$ is the strength of the time periodic forcing,
$A$ and $\phi_a$ are parameters that allow to introduce a spatial asymmetry,
and $B$ and $\phi_b$ have an analogous effect in the time domain.
Throughout this paper we will take $A=1/2$ and $\phi_b=\pi/2$.
The effects of the environment are introduced
by means of a velocity dependent damping and thermal fluctuations.
This leads to integrate the equation
%
\begin{equation}
  m \ddot{x}=-\Gamma \dot{x} - V'(x,t) + \xi.
\end{equation}
In the above expressions $x$ is the spatial coordinate
of the particle, $m$ its mass, and $\Gamma$, with $ 0 \leq \Gamma \leq 1$,
 is the dissipation parameter.
The thermal noise $\xi$ is related to $\Gamma$, according to
$ <\xi^2> =2 (1-\Gamma) k_B T$, where $k_B$ is the Boltzmann constant and $T$
is the temperature, thus making the formulation consistent with the
fluctuation-dissipation relationship.
In the following we take $m=1$ and $k_B=1$.

At the quantum level, we perform the evolution of the density matrix of
the system, $\rho$, by means of a modified split operator method \cite{SplitOp}.
We use a composition of unitary steps given by the kinetic and potential
terms of the Hamiltonian (representing the system dynamics),
and other purely dissipative steps.
The latter come as the result of incorporating dissipation and thermalization
to the quantum particle by coupling it to a bath of noninteracting oscillators
in thermal equilibrium at a temperature $T$.
The degrees of freedom of the bath are eliminated by means of
the usual weak coupling, Markov and rotating wave approximations \cite{QNoise}.
As a result, we arrive at a Lindblad equation for the density matrix of the system,
that can be written as a completely positive map ${\bf D}_\alpha(dt)$ in the
operator-sum (or Kraus) representation
%
\begin{equation}
  \rho(t+dt) = {\bf D}_{(\varepsilon,T)}(dt)\left(\rho(t)\right)=
               \sum_{\mu=0}^{2}{K^\pm_\mu \rho(t) K^{\pm \dagger}_\mu},
  \label{krausrep}
\end{equation}
where
%
\begin{eqnarray}
  K_0     & = & \mathbbm{1}-\frac{1}{2} \;
                \sum_{\mu=1}^{2}{K^{\pm \dagger}_\mu K^\pm_\mu} \nonumber \\
  K_1^\pm & = & \sum_{k=1}^{N-1} \sqrt{\varepsilon \;dt \; (1+\bar{n}(k)) \; k}
                \; \ket{p_{\pm k \mp 1}}\bra{p_{\pm k}}  \nonumber \\
  K_2^\pm & = & \sum_{k=1}^{N-1} \sqrt{\varepsilon \;dt \; \bar{n}(k) \; k}
                \; \ket{p_{\pm k}}\bra{p_{\pm k \mp 1}}
  \label{krauslindblad}
\end{eqnarray}
are the infinitesimal Kraus operators satisfying:
$\sum_\mu{K^{\pm \dagger}_\mu K^\pm_\mu}=\mathbbm{1}$ to first order in $dt$ \cite{Carlo},
and $p$ is the momentum conjugated to the $x$ coordinate.
Note that the superscript $\pm$ defines two different operators (standing for the positive
and negative values of the $p$ spectrum), and do not apply to the $K_0$ operator.
In these equations $\varepsilon$ is a system-bath coupling parameter that can be
directly associated to the classical velocity dependent damping $\Gamma$
(at $T=0$ gives the contraction rate of the phase space).
The population densities of the bath are given by
$\bar{n} = (\exp(\Delta E_{k}/(k_B T))-1)^{-1}$, where we have taken $E_{k} = p_{k}^2/2$,
and $\Delta E_{k}$ is the energy difference between the neighboring levels connected
by the operators.
In this way, we extend our method  \cite{ThEff}, originally developed for maps,
to general fluxes.

\section{Results}
  \label{sec:results}

In directed transport studies the main quantity that characterizes the system
is the current $J(t)=\langle p_t \rangle$, where $\langle \rangle$ means the
average taken with respect to the initial conditions and time, up to a given instant $t$.
In the quantum case, we consider $J(t)=\langle \tr{(\rho p)} \rangle$,
where the same kind of average is taken.
We study its transitory and stationary behavior.
Given that phase space distributions are also of great interest,
in the following we will also focus on their analysis.

We have considered two cases, weak and strong forcing.
In the first case ($F \sim 0.02 - 0.05 $) we have broken all spatiotemporal
symmetries that forbid a net current
(namely, a generalized parity and time reversal) by means of the potential
built-in asymmetry (we have taken $B=0.2$ and $\phi_a=0$).
In this situation the phase space of the closed system is mixed and the
current depends on the structure of the chaotic sea, which has to be asymmetric.
Moreover, we have tested the robustness of this mechanism
against dissipation and thermal fluctuations.
In the second case ($F=2.5$) we have chosen a simpler, harmonic forcing ($B=0$),
and an asymmetric spatial potential ($\phi_a=\pi/2$) in order to investigate
the interplay among forcing, dissipation, and thermal fluctuations.
Now, the current comes from the asymmetry of a strange attractor.
In fact, dissipation induces this asymmetry, which is responsible for the
directed transport \cite{QChDissRat}.
But this same dissipation mechanism contracts phase space and makes the higher
energies inaccessible for the system.
So higher dissipation and asymmetry does not translate necessarily 
into higher values of $J$.
Additionally, thermal noise compensates for the energy loss caused by dissipation.
However, the associated diffusion tends to homogenize the strange
attractor, reducing its asymmetry, and these two effects compete
with each other \cite{ThEff}.

\subsection{Weak forcing}

\begin{figure}[htp]
\begin{center}
\vspace{0.05\textwidth}
  {\epsfxsize=8cm\epsffile{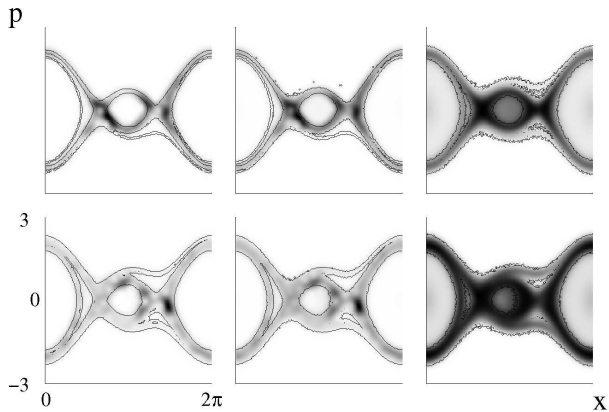}}
  \caption{Classical (contour lines) and quantum (gray scale density plot) phase space
      distributions for $t=50$ (in units of the period of the forcing).
      In the upper row $F=0.02$ and in the lower one $F=0.05$.
      We display the cases (from left to right) for $T=0$, $T=0.01$, and $T=0.1$.
      In all cases $\Gamma=10^{-4}$ and $\hbar=0.041$.}
\label{fig:wfphasespace}
\end{center}
\end{figure}

The biharmonic spatial potential we have considered has two minima.
For weak time-periodic forcing there is a chaotic region that connects
two corresponding main regular islands which are related to these minima.
In fact, a chaotic region fills the portion of phase space among many regular islands,
in the shape of a network of branches, with almost no bulk portion.
As a consequence, there is a high sensitivity to variations in the intensity
of the driving field, since this introduces relevant modifications in this
kind of ballistic network.
This phenomenon can be clearly seen in Fig.~\ref{fig:wfphasespace},
which displays the phase space for $F=0.02$ (upper row) and $F=0.05$ (lower row)
at time $t=50$.
In all cases the coupling with the environment is very weak, $\Gamma=10^{-4}$.
In addition we take $\hbar=0.041$, which is far from the semiclassical limit.
The classical and quantum distributions are obtained by evolving a set of 
analogous initial conditions inside the chaotic region with initial $p=0$.
The contour lines indicate different level surfaces of the classical distributions
while the density plots show the quantum ones (where darker means higher values).
%
The imbalance between the positive and negative $p$ regions of phase space,
which is responsible for the current, is significantly altered by varying the parameter $F$.
For $F=0.02$ and $T=0$ (see Fig. \ref{fig:wfphasespace}, upper row, leftmost panel)
there is a branch of the chaotic region that develops for $p < 0$ and that is not
present for $p>0$. For $F=0.05$ and $T=0$ (see Fig.~\ref{fig:wfphasespace}, lower row,
leftmost panel) the distribution looks more symmetric.
This behavior persists at low temperatures ($T=0.01$) as can be seen in the middle column of
Fig.~\ref{fig:wfphasespace}, for both values of $F$.
Finally, for a higher temperature ($T=0.1$) the situation changes, and the distributions
look more alike, although some features of the phase space structure seem to survive
the thermal fluctuations.

\begin{figure}[htp]
\begin{center}
\vspace{0.05\textwidth}
 {\epsfxsize=8cm\epsffile{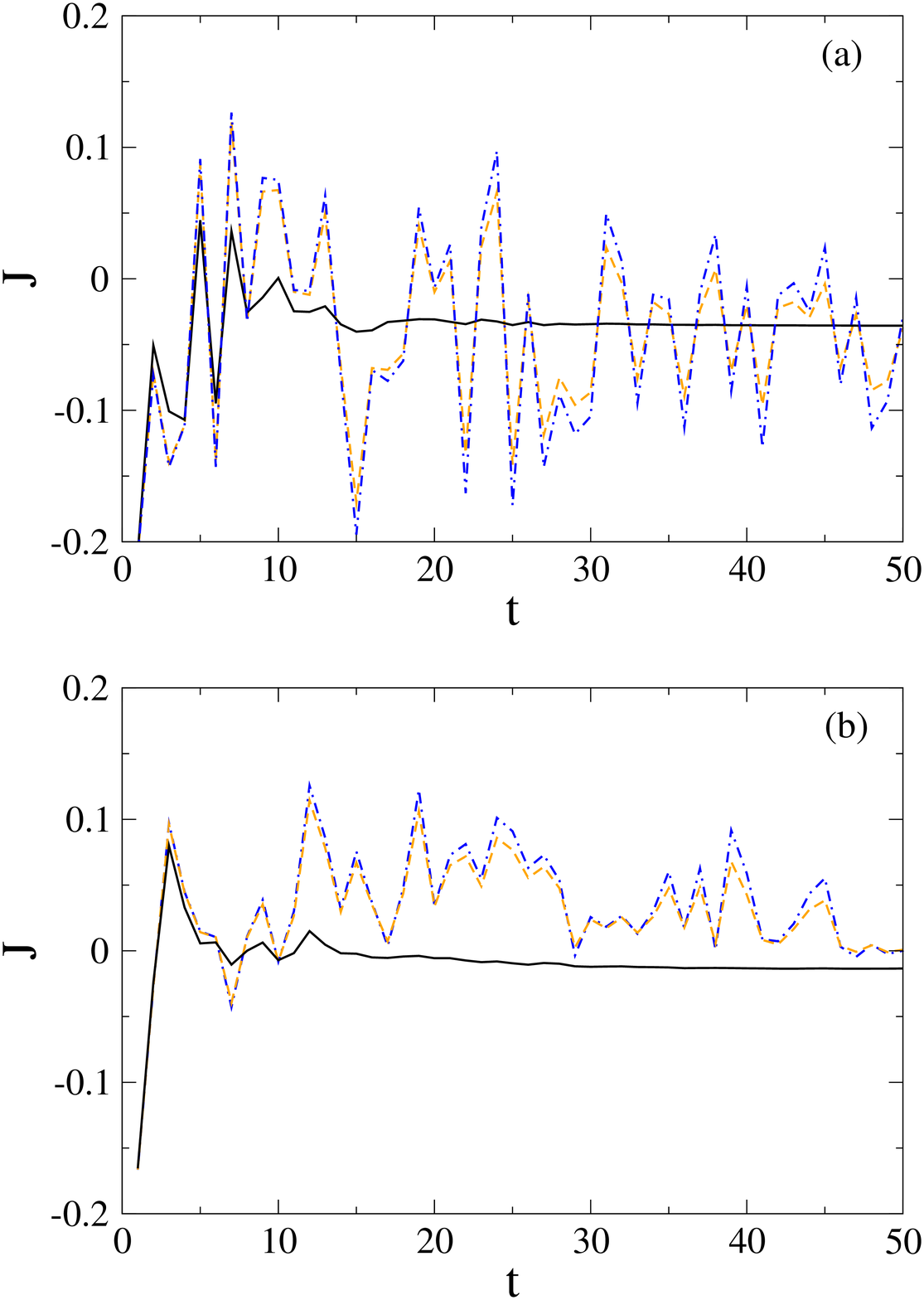}}
  \caption{(Color online) Quantum current as a function of time $t$
    (in units of the period of the forcing) for the cases shown in
    Fig.~\ref{fig:wfphasespace}.
    In panel (a) $F=0.02$ and in (b) $F=0.05$.
    (Blue) dot-dashed lines correspond to $T=0$,
    (orange)   dashed lines to $T=0.01$, and
    (black) solid lines to $T=0.1$.
    In all cases $\Gamma=10^{-4}$ and $\hbar=0.041$.}
 \label{fig:wfcurrents}
\end{center}
\end{figure}

In order to assert if this is the case, we show the current values as a function
of time (in units of the period of the forcing) in Figs.~\ref{fig:wfcurrents}
(a) (for $F=0.02$) and (b) (for $F=0.05$).
%
We notice that at $T=0$ (blue dot-dashed lines) there are strong fluctuations that can
be associated with the very weak coupling with the environment and the finite $\hbar$ value.
At $T=0.01$ (orange dashed lines) the same behavior is present,
but at $T=0.1$ (black solid lines) the situation changes and the current stabilizes
at very short times ($t \sim 15$), so that a value for the current
$J$ can be defined. Moreover, this is valid for both values of $F$. 
At this point it is worth mentioning that for $F=0.02$ the current
is greater than for $F=0.05$, so we can conclude that the quantum distributions
preserve the main features of the original structure of phase space at $T=0$.
This makes the current generation mechanism robust against environmental perturbations
in the weak coupling regime. Moreover, temperature also helps to stabilize the ratchet.

To conclude this subsection, we would like to mention that
our choice of parameters is suitable for modeling many isomerization 
reactions induced by laser fields \cite{isovision,isocontrol,isomerization}. 
The interactions with the solvent can account for the interaction with a thermal
environment of the kind that we have analyzed. In this respect, it is interesting to note 
that the main features for current generation (isomerization rate in this case)
survive a weak coupling with the environment, and that this in turn can be beneficial.
This suggests that this mechanism is applicable to actual experimental situations.

\subsection{Strong forcing}

\begin{figure}[htp]
\begin{center}
  \vspace{0.05\textwidth}
 {\epsfxsize=8cm\epsffile{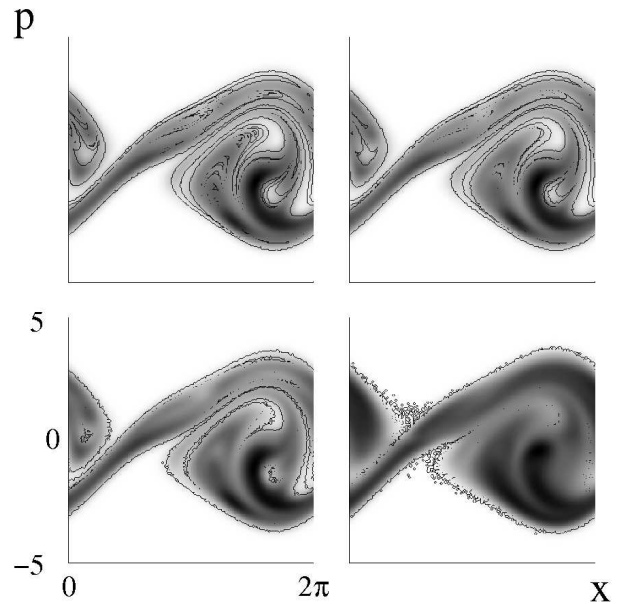}}
  \caption{Classical (contour lines) and quantum (gray scale density plot) phase
  space distributions for $t=50$ (in units of the period of the forcing).
  From left to right and top to bottom we show
  the results corresponding to: $T=0$, $T=0.001$, $T=0.01$ and $T=0.1$, respectively.
  In all cases $F=2.5$, $\Gamma=0.05$ and $\hbar=0.041$.}
\label{fig:sfphasespace}
\end{center}
\end{figure}

\begin{figure}[htp]
\begin{center}
  \vspace{0.05\textwidth}
 {\epsfxsize=8cm\epsffile{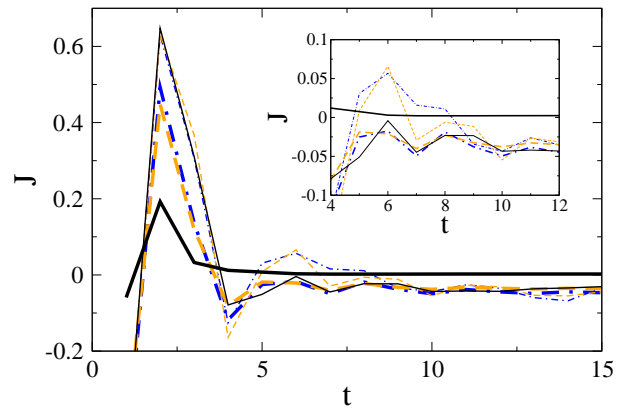}}
  \caption{(Color online) Classical (thin lines) and quantum current as a function
  of time $t$ (in units of the period of the forcing) for some of the cases
  shown in Fig.~\ref{fig:sfphasespace}.
  (Blue) dot-dashed lines correspond to $T=0$,
  (orange) dashed lines to $T=0.01$, and
  (black) solid lines to $T=0.1$.
  In all cases $F=2.5$, $\Gamma=0.05$ and $\hbar=0.041$.
  In the inset we show enlarged the stabilization region for the sake
  of clarity.}
\label{fig:sfcurrents}
\end{center}
\end{figure}

The same system considered above can behave in a quite different way if the strength of
the forcing is increased. In fact, when this happens the regular islands structure present
in the previous case is completely lost, and then the current arises as a consequence
of the asymmetry of a chaotic attractor.
In this case, it is easier to define an asymptotic current given the fact 
that the attractor is usually formed in a very short time.
To illustrate this effect, we present in Fig.~\ref{fig:sfphasespace} 
the classical and quantum phase space distributions for $F=2.5$, $\Gamma=0.05$, and
$\hbar=0.041$ at $t=50$.
In it, we show from left to right and top to bottom the cases corresponding to:
$T=0$, $T=0.001$, $T=0.01$ and $T=0.1$, respectively.
%
As in the previous case, the influence of the temperature
is noticeable only for its higher value, namely $T=0.1$; while for the rest of 
the cases the quantum distribution remains almost unchanged. Thermal fluctuations 
need to be greater than the quantum ones to manifest themselves.
This is not the case for the classical attractor,
which gradually loses all its finer details.
For the set of parameters that we have chosen in this case, the time reversal invariance
is only broken by dissipation.
Moreover, dissipation is not only responsible for the current generation,
but also for the quick stabilization of the current.
This can be seen in Fig.~\ref{fig:sfcurrents}, where the classical (thin lines) and
quantum $J$ as a function of time $t$ are shown.
%
As can be seen, for $T=0$ and $T=0.01$ the attractor needs $\sim 10$
periods of the forcing to set in;
at this point the current is very well defined, both classical and quantum mechanically.
Actually, the correspondence between the quantum and classical results is remarkable. 
For $T=0.1$ the setting in of the attractor is even faster
(just $\sim 6$ periods are needed), but in this case 
the effect of temperature on the quantum current is rather drastic, i.e.~$J$ goes to zero.
We have further verified this behavior by using a highly efficient
diagonalization method \cite{futurework}.
This allowed us to look into the details of the superoperator spectrum,
for which the spectral gap is around $0.4$ (this gap is the distance between the modulus 
of the leading eigenvalue - equal to one - and the modulus corresponding to the next eigenvalue, 
in descending order of this quantity).
We have also obtained the associated equilibrium eigenvector (corresponding to the eigenvalue one), 
which resulted indistinguishable from the chaotic attractor found
by means of the time evolution.

\begin{figure}[htp]
\begin{center}
  \vspace{0.05\textwidth}
 {\epsfxsize=8cm\epsffile{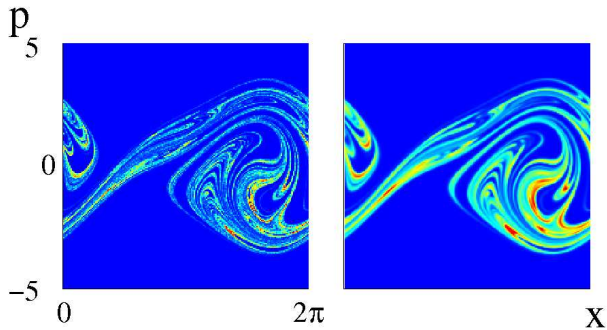}}
  \caption{(Color online) Classical (left panel) and quantum (right panel)
  phase space distributions at time $t=50$ corresponding to $T=0$.
  In all cases $F=2.5$, $\Gamma=0.05$ and $\hbar \simeq 0.0015$.}
 \label{fig:sfscphasespace}
\end{center}
\end{figure}

In order to show the ability of our method to reach the semiclassical limit we consider 
the classical and quantum chaotic attractors corresponding to $\hbar \simeq 0.0015$ and $T=0$
in Fig.~\ref{fig:sfscphasespace}. Let us mention here that the study of the evolution of different quantities
as a function of time is an interesting sub-product of our approach.
Thanks to this, we have found that despite the greater accuracy in the details
of the phase space distributions attained in this case, the values of $J$ as a function of $t$ remain
essentially the same as for $\hbar=0.041$.
%

\section{Conclusions}
  \label{sec:summary}

In this paper, we have studied a ratchet system consisting of a particle moving
in a biharmonic potential subjected to a time periodic forcing and
in contact with a heat bath at finite temperature.
The rich structure of the phase space for different values of the
involved parameters has been analyzed both classical and quantum mechanically.

We have found that for weak forcing and very weak coupling with the environment,
the mechanism leading to current generation survives.
In fact, the quantum distributions keep ``memory'' of the original structure
of the chaotic region at $T=0$.
Moreover the effect of the temperature is beneficial, rather than destructive,
and the current stabilizes after short times.
This makes us suggest a possible experimental verification of these results 
in chemical isomerization reactions,
to which the dynamics we have analyzed can be directly applicable.
In the case of the strong forcing regime, the effect of temperature 
also shorten the times for which the stabilization of the current occurs.
But in this case, the current 
vanishes for higher values of $T$, due to the blurring of the chaotic
attractor details (and asymmetry).

Finally, it is worth mentioning the high precision and simplicity
of our integration method.
By developing a modified split operator scheme, we have been 
able to reach very low values of $\hbar$. 
This, combined with a highly efficient diagonalization method
will allow us to implement in the near future an alternative approach 
for studying time dependent dissipative quantum systems \cite{futurework}.

\begin{acknowledgments}
Partial support by ANPCyT and CONICET (Argentina), and to
MICINN--Spain, under contracts No.~MTM2009--14621 and
iMath--CONSOLIDER 2006--32, is gratefully acknowledged.
\end{acknowledgments}

\end{document}